# Navigated hepatic tumor resection using intraoperative ultrasound imaging

Karin A. Olthof[1,2,*] Theo J.M. Ruers[1,2] MD, PhD, MSc, Tiziano Natali[1,2] MSc, Lisanne P.J. Venix[1,2] MSc, Jasper N. Smit[1] PhD, Anne G. den Hartog[1] MD, PhD, Niels F.M. Kok[1] MD, PhD, Matteo Fusaglia[1] PhD, Koert F.D. Kuhlmann[1] MD, PhD

[1] Department of Surgical Oncology, Netherlands Cancer Institute, Plesmanlaan 121, 1066CX, Amsterdam, The Netherlands

[2] Faculty of Science and Technology (TNW), Nanobiophysics Group (NBP), University of Twente, Drienerlolaan 5, 7522 NB, Enschede, The Netherlands

* Corresponding author: ka.olthof@nki.nl, +3120 512 2531

**Funding:** This research received no external funding.

**Category:** Original article

**Conflicts of Interest:** The authors (KO, TN, LV, JS, AH, NK, TR, MF, KK) declare no conflict of interest.

**Data Availability Statement:** The datasets generated during and/or analyzed during the current study are not publicly available.

**Ethical Statement:** All procedures performed in this study involving human participants were conducted in accordance with the ethical standards of the institutional and national research committee.

**Abstract**

*Purpose:* This proof-of-concept study evaluates feasibility and accuracy of an ultrasound-based navigation system for open liver surgery. Unlike most conventional systems that rely on registration to preoperative imaging, the proposed system provides navigation-guided resection using 3D models generated from intraoperative ultrasound.

*Methods:* A pilot study was conducted in 25 patients undergoing resection of liver metastases. The first five cases served to optimize the workflow. Intraoperatively, an electromagnetic sensor compensated for organ motion, after which an ultrasound volume was acquired. Vasculature was segmented automatically and tumors semi-automatically using region-growing ($n$=15) or a deep learning algorithm ($n$=5). The resulting 3D model was visualized alongside tracked surgical instruments. Accuracy was assessed by comparing the distance between surgical clips and tumors in the navigation software with the same distance on a postoperative CT of the resected specimen.

*Results:* Navigation was successfully established in all 20 patients. However, four cases were excluded from accuracy assessment due to intraoperative sensor detachment (n=3) or incorrect data recording (n=1). The complete navigation workflow was operational within 5-10 minutes. In 16 evaluable patients, 78 clip-to-tumor distances were analyzed. The median navigation accuracy was 3.2 mm [IQR: 2.8–4.8 mm], and an R0 resection was achieved in 15/16 (93.8%) patients and one patient had an R1 vascular resection.

*Conclusion:* Navigation based solely on intra-operative ultrasound is feasible and accurate for liver surgery. This registration-free approach paves the way for simpler and more accurate image guidance systems.

**Introduction**

The primary objective of liver surgery is to achieve complete tumor resection, while sparing parenchyma and preserving critical vascular and biliary structures. Intraoperatively, ultrasound imaging is the primary modality to localize tumor borders and their spatial relationships to intrahepatic vasculature. Nevertheless, two-dimensional (2D) ultrasound can be complex to interpret, is operator-dependent and lacks continuous feedback during resection. Furthermore, once resection has started, gas bubbles generated by electrocautery devices disrupt the ultrasound signal, complicating its interpretation. These limitations contribute to a relatively high incidence of R1 resections, reported in 14–22% of patients (1,2).

Image-guided surgical navigation addresses these challenges, offering a live, virtual representation of the surgical scene, using a three-dimensional (3D) model of the organ that includes tumors and critical structures. The technique has shown its value in intraoperative localization of small and vanished hepatic lesions (3–5). While several studies investigated its application in hepatic resection, it has not yet demonstrated sufficient accuracy or clinical benefit to surpass conventional methods (6–10). A major limitation of conventional navigation systems is their reliance on registered preoperative 3D models, which in liver surgery are often unreliable due to changes in the shape of the deformable organ during the surgical procedure.

Non-rigid registration algorithms have been explored to account for organ deformation and improve the accuracy of liver navigation systems (11–14). However, these methods are generally computationally demanding, introduce additional complexity during surgery, and cannot continuously compensate for further deformations during resection. Registration between preoperative imaging and the intraoperative patient anatomy therefore remains the main source of error in surgical liver navigation. Moreover, if tumor progression or ongoing systemic therapy response occurs between the preoperative scan and surgery, the navigation system remains inherently inaccurate.

For small or vanishing lesions that cannot be identified with ultrasound alone, navigation requires registration to a preoperatively segmented 3D model. In contrast, for larger lesions that are visible on ultrasound, navigation systems can base their information solely on this intraoperative imaging modality (15–19). As these systems eliminate the need for registration, navigation can be performed more accurately. Potentially, navigation then aids a more accurate resection. Nonetheless, current literature on ultrasound-only liver navigation remains limited, and existing approaches have not demonstrated tumor segmentation from ultrasound volumes.

We describe a novel electromagnetic (EM) image-guidance system that generates three-dimensional liver models directly from intraoperative ultrasound to assist surgical resections. This study evaluates the feasibility and accuracy of the proposed navigation system.

**Methods**

*Patient population*
A prospective, single center feasibility study was conducted at the Netherlands Cancer Institute. Between December 2022 and February 2025, 25 patients 18 years and older were selected for ultrasound-based navigated surgical resection. Selection criteria were patients with tumors >2 cm, visible on intraoperative ultrasound. Exclusion criteria were patients scheduled for standard-of-care navigation for localization of vanished liver lesions (3), those scheduled for anatomical hemi-hepatectomy and patients with a pacemaker, due to the potential interference of the electromagnetic field generator. In patients with multiple tumors, navigation was performed during resection of a single lesion. Navigation performed in the first five patients served to optimize the surgical workflow. Data of these patients was not included in the accuracy analysis. This study was approved by the institutional review board (NL80634.031.22). All patients provided written informed consent.

*Tracked instruments*

Tracked surgical instruments included an US transducer (type I14C5T, BK Medical, Denmark), a vessel sealer (LigaSure Impact™, Medtronic, MN, USA) and a surgical pointer (Aurora 6DOF Probe, NDI) (Figure 1). The US transducer and vessel sealer were equipped with EM sensors using a custom 3D-printed clip (Nylon PA12) and a serializable stainless steel (AISI 316) adapter, respectively. The adapter for the vessel sealer was designed to rotate synchronously with the instrument's shaft, allowing adequate visualization of its orientation. Both adapters were calibrated prior to the procedure. Real-time tracking of the surgical instruments was achieved by placing the Aurora® V2 planar field generator (Northern Digital Inc., Waterloo, Ontario, Canada) near the surgical field.

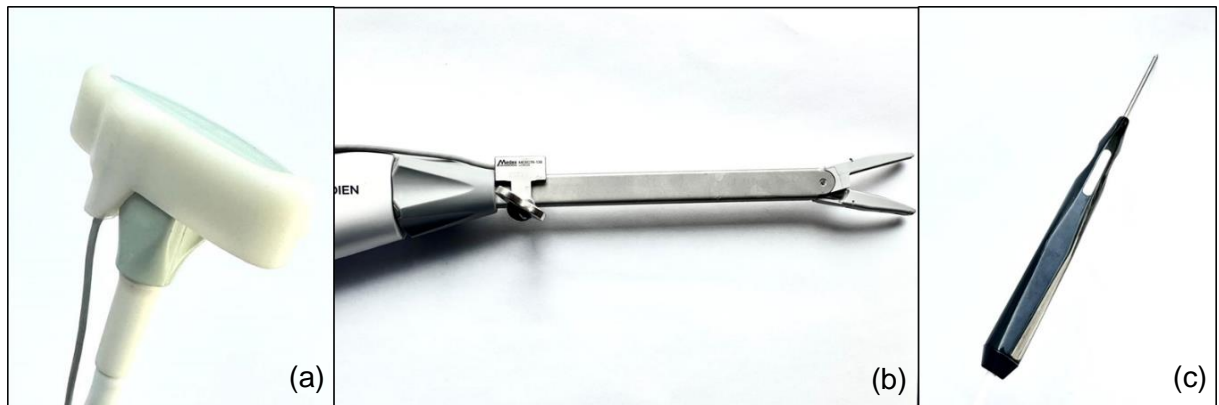

(a) (b) (c)

**Figure 1:** EM tracked surgical instruments used in this study: (a) intraoperative ultrasound transducer, (b) vessel sealer, (c) surgical pointer

*Surgical navigation workflow*

Intraoperatively, the liver was first mobilized to ensure adequate exposure of the tumor. To account for tumor movement caused by respiration and surgical manipulation, an EM sensor was temporarily glued to the parenchyma in close proximity of the tumor (Dermabond® advanced adhesive, Ethicon), as previously described in (10). Surgical navigation was provided using a custom module created in 3D Slicer (20). PlusServer, part of the PLUS toolkit (21), was used to obtain real-time tracking and imaging data and sent to 3D Slicer via the OpenIGTLink module.

To include spatial understanding of the US-based model in relation to the entire liver, the parenchyma and the target lesion were preoperatively delineated from the most recent diagnostic imaging, which was either contrast-enhanced computed tomography (CT) or magnetic resonance imaging (MRI). A single-point registration was then performed, by aligning the tumor's center found with US with the corresponding center derived from preoperative imaging (22).

Subsequently, an intraoperative volume of the tumor and surrounding vasculature was acquired by sweeping the tracked US transducer over the liver surface. Vasculature was then segmented from the US volume, using a deep learning algorithm based on the 3D U-Net architecture (23), trained to segment vessels from reconstructed hepatic US volumes. Tumor segmentation was performed semi-automatically by a technical physician in the operating room using one of two methods. In the first twenty patients, tumor segmentation was performed using a region growing algorithm in 3D Slicer, by placing seed points within and outside of the tumor. In the last five patients, a deep learning approach was used. A region of interest around the tumor was defined, serving as input for the automated segmentation framework, as described in a publicly available preprint.(24) For both segmentation approaches, delineations were assessed and confirmed by the operating surgeon. When segmentation was not satisfactory, adjustments were performed manually using the Segment Editor module in 3D Slicer. Based on the confirmed tumor boundaries, a virtual resection margin was added of either 5, 7, or 10 millimeters and visualized.

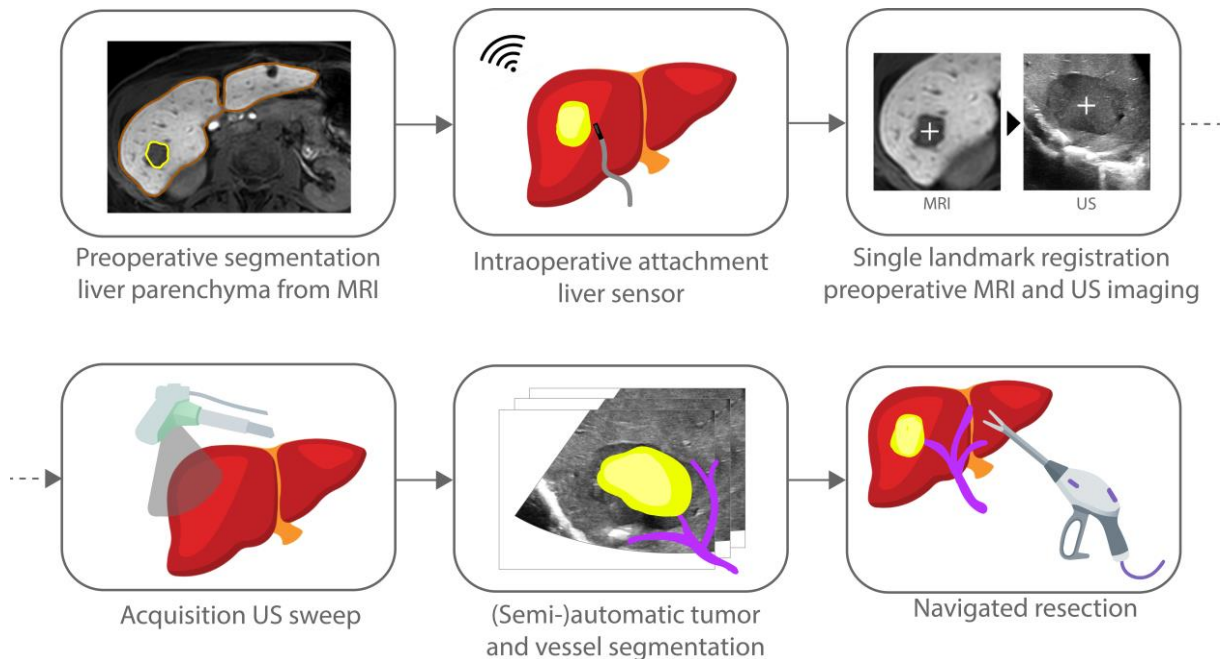

**Figure 2:** Schematic overview of study workflow.

*Visualization*
To support intraoperative orientation, a dual-view of the virtual scene was implemented, based on the surgeon's feedback. Two screen layouts were selected as standard. The first layout displayed the tracked US probe along with the US image plane overlaid with the segmented tumor model, alongside the 3D view (Figure 3a). This enabled immediate verification of tumor and vessel segmentation accuracy and allowed the planned resection plane to be directly visualized in the intraoperative US image. The second screen-layout was used during resection and included two orthogonal 3D perspectives of the segmented anatomy: a cranial (top-down) and a lateral (side) view (Figure 3b). In addition, the shortest distance from the surgical instrument (i.e., the vessel sealer or the tracked pointer) to the tumor border was displayed. If required, the 3D model could be interactively rotated to align the visualization with the surgeon's perspective or the intraoperative imaging.

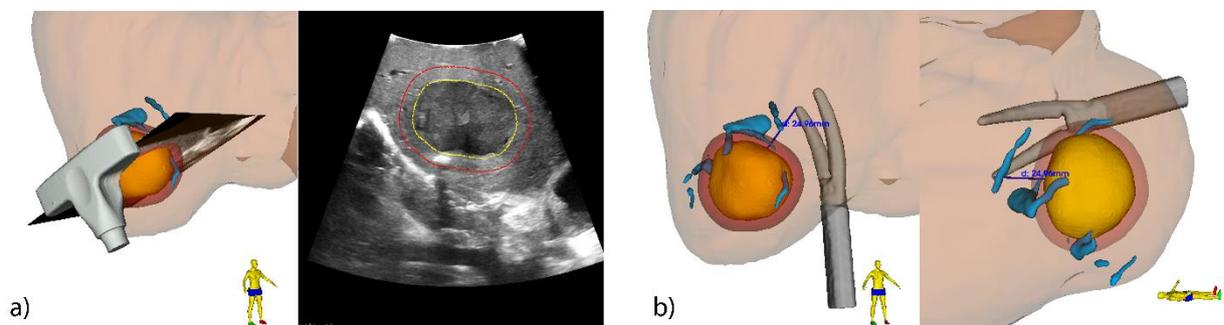

**Figure 3:** Two dual-view visualizations of surgical navigation: a) 3D view showing the tracked US probe with respect to the segmented tumor (yellow), proposed resection margin (red) and vasculature (blue), and a cross-sectional overlay of the segmented 3D model over the live US image. b) Two 3D views from top and side view of the patient, showing the tracked vessel sealer with the shortest distance to the tumor border.

*Evaluation*

We documented the time of workflow steps (single landmark registration, ultrasound volume acquisition and reconstruction and tumor segmentation), the navigation accuracy and the histopathologic findings of the surgical specimen. The navigation accuracy was assessed by placing surgical clips along the resection plane during resection. The number of clips depended on the tumor shape and size. The shortest distances from the clips to the tumor border in the navigation software were recorded with the tracked pointer (Figure 4a). Postoperatively, CT scans of the resected specimens were acquired with a slice thickness of 0.5 mm. Tumors and clips were manually segmented, and the shortest 3D distances between clips and tumor margins were calculated (Figure 4b). These postoperative measurements were compared to the intraoperative values to quantify navigation accuracy. Navigation accuracy was assessed both per clip and per patient, where per-patient accuracy was computed as the average value of all clips placed during a single procedure. All analyses were performed in Python (version 3.9, Python Software Foundation, https://www.python.org). In addition, qualitative feedback on system usability was collected from the operating surgeons during and immediately after the procedures and included observations on visualization, workflow integration, and perceived usefulness during different phases of resection.

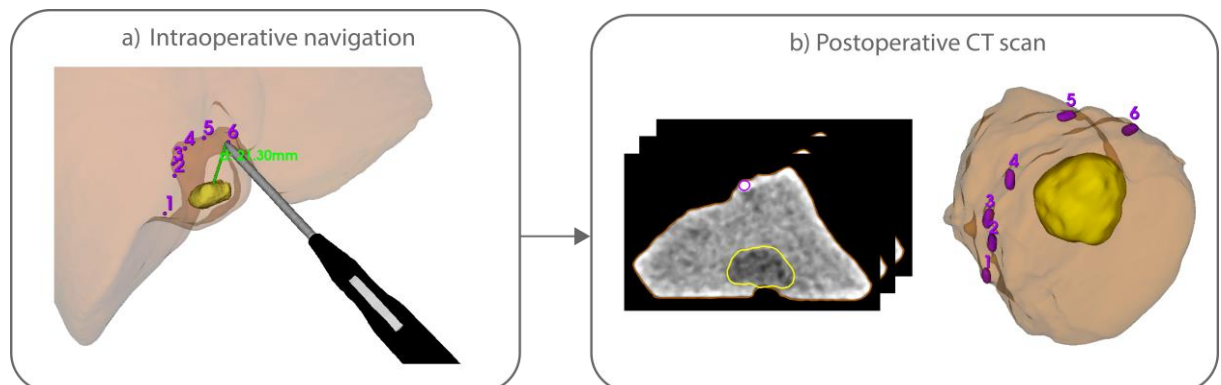

**Figure 4:** Intraoperative and postoperative assessment of distances from clips (purple) to tumor (yellow) for evaluating navigation accuracy. a) During surgery, surgical clips are placed along the resection plane. Their positions were digitized using a tracked pointer, allowing intraoperative assessment of resection margins by the navigation software. b) After resection, a postoperative CT scan of the specimen was acquired. The distances from each clip to the tumor boundary were measured and compared to the intraoperative measurements. This comparison enabled quantification of the accuracy of the navigation system.

**Results**

Patient characteristics of the 20 patients included in the study are shown in Table 1.

**Table 1:** Patient characteristics

| Characteristic | n=20, n (%) or median [range] |
|---|---|
| Sex | |
|     Male | 11 (55.0) |
|     Female | 9 (45.0) |
| Age at surgery (years) | 63 [37 – 82] |
| Neoadjuvant treatment | |
|     None | 5 (20.0) |
|     Systemic chemotherapy | 14 (76.0) |
|     HAIP combined with systemic chemotherapy | 1 (4.0) |
| Tumor type | |
|     CRLM | 19 (95.0) |
|     GIST | 1 (5.0) |
| Number of tumors | 2 [1 – 12] |
| Tumor diameter (mm) | 31 [20 – 73] |
| Tumor segment |  |
|     II | 1 (5.0) |
|     III | 2 (10.0) |
|     IVa | 3 (15.0) |
|     IVb | 2 (10.0) |
|     V | 1 (5.0) |
|     VI | 3 (10.0) |
|     VII | 7 (30.0) |
|     VIII | 1 (5.0) |

*HAIP* hepatic arterial infusion pump, *CRLM* colorectal liver metastasis, *GIST* gastrointestinal stromal tumor

*Feasibility*
One technical physician was required in the operating room to set up the system and operate software. In all patients, navigation was performed. The liver sensor that was glued to the liver surface detached in 3/20 (15.0%) of patients during resection due to tissue handling by the surgeon. This results in loss of navigation, since the coordinate system was defined relative to the reference sensor. In one patient, data was incorrectly saved, leaving 16 patients for accuracy evaluation. Single landmark registration and ultrasound acquisition/reconstruction took an average of 45 [range: 10 – 225] and 67 [range: 21 – 136] seconds, respectively. Semi-automatic tumor segmentation by region-growing and using a deep-learning network took an average of 323 [range: 122 – 780] and 269 [range: 82 – 600] seconds. The use of the deep-learning approach for segmentation reduced segmentation time by approximately one minute, although manual refinement was still required. A complete overview of the navigation workflow, including deep-learning based automated tumor segmentation, is demonstrated in the **Video**.

*Navigation accuracy*
In the 16 patients included in the accuracy analysis, a total of 84 surgical clips was placed, with a median of 5 clips [range 4–6] per patient. Of these, 6 clips (7.1%) detached during handling or positioning in the pathology tray. Navigation accuracy was assessed both per clip (n = 78) and per patient (n = 16), where per-patient accuracy was computed as the average value of all clips placed during a single procedure (Figure 5). The median navigation accuracy per clip was 3.3 mm (IQR: 2.0–5.3 mm). Per patient, the median accuracy was 3.2 mm (IQR: 2.8–4.8 mm). One clear outlier was observed, corresponding to a patient with a large tumor of 6 cm in diameter in segments VI and VII. In this case, the reference sensor had been positioned on the lateral surface of the liver, on the opposite side of the tumor relative to the resection plane. As a result, the sensor did not adequately track local tumor motion relative to the resection area, resulting in the outlier observed in accuracy. An R0 resection was obtained in 15/16 (93.8%) of the procedures. In one patient, an R1 vascular resection was deliberately performed in order to preserve the right hepatic vein; microwave ablation was applied at the resection plane to treat residual tumor.

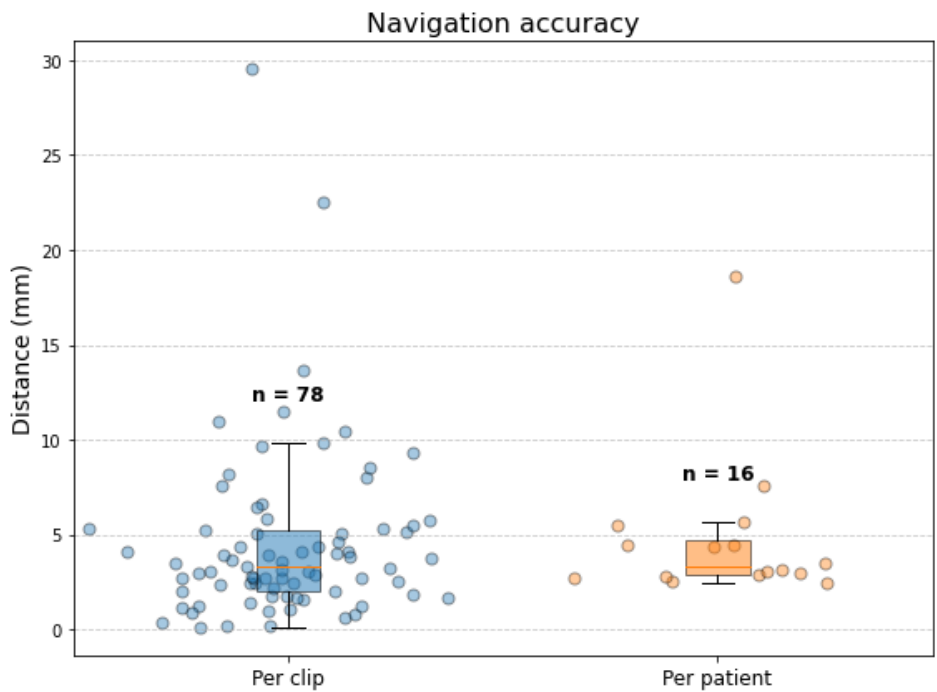

**Figure 5:** Boxplots showing the distribution of navigation accuracy distances measured per clip and averaged per patient. Boxes represent the median and interquartile range; whiskers indicate the data range excluding outliers. Individual data points are overlaid.

*Usability*
Surgeons reported that the ultrasound overlay was useful for verifying segmentation accuracy. Although visualization of both cranial and lateral perspectives of the 3D scene intended to offer a comprehensive overview of the segmented anatomy, surgeons reported difficulty in determining the precise location of surgical instruments in relation to the tumor. This was mainly attributed to limited depth perception when interpreting 3D structures on a 2D screen. Spatial interpretation improved when the view was manually adjusted to align with the surgeon's perspective. The display was regarded as particularly useful for verifying resection margins in cases of uncertainty, with the measured distance between the surgical instrument and tumor margin identified as the most valuable feature. However, during critical phases of resection, continuous reference to the screen was considered impractical. To address this limitation, an auditory alert was introduced to notify the surgical team when an instrument approached the planned tumor margin. For small, peripheral resections, surgeons noted that the additional time required for navigation outweighed its potential benefits, and the system was not considered essential in such cases.

**Discussion**

This is the first study that investigated an ultrasound-based guidance system that provides continuous feedback during open resections of liver tumors. Unlike conventional navigation systems that rely on preoperative imaging and are susceptible to anatomical changes and registration errors, the proposed approach uses intraoperative ultrasound data to generate a 3D model of the tumor and vasculature and allows for tracking of the tumor during the resection. The proposed system enabled accurate instrument guidance, achieving a median navigation accuracy of 3.2 mm in the resection plane. To the best of our knowledge, this represents the highest navigation accuracy reported to date for open liver surgery.

Comparable to this study, Beller et al. (15) evaluated the use of an optically tracked 3D US probe and cavitron ultrasonic surgical aspirator (CUSA) to guide liver resections. The system was successfully applied in 52 of 54 patients, in which surgeons used two orthogonal ultrasound section planes and for visualization. While vasculature could be extracted semi-automatically, fast tumor segmentation was not achieved. Histological analysis revealed a mean resection margin of 9 mm, with maximal deviation of 8 mm from the pre-planned margins. In contrast, Paolucci et al. (17) incorporated tumor information into the navigation system through semi-automatic 2D segmentation on a central ultrasound slice. The resulting contour was used to generate a spherical approximation of the tumor volume and to automatically compute a resection strategy. This approach yielded R0 resections in 22 out of 23 cases in an ex vivo setting, demonstrating feasibility of the method. The assumption of a spherical tumor shape, however, limits the applicability of this method, particularly in cases involving large or irregularly shaped tumors.

In open liver surgery, resection relies on preoperative imaging, palpation, and intraoperative ultrasound. Ultrasound-based navigation adds value in this context by providing continuous, 3D information on tumor boundaries and their relation to critical structures. In contrast, during minimally invasive procedures, near-infrared fluorescence-guidance with the fluorescent dye indocyanine green (ICG) forms an alternative to aid in lesion differentiation and guidance of surgical margins, and is often more accurate than navigation in this context.(25) Instead, in minimally invasive surgery, navigation mainly supports the localization of small or deep lesions and the visualization of critical anatomical structures such as vasculature and bile ducts, typically achieved by registration of preoperative model.

However, fluorescence-based guidance is limited to superficial lesions due to the restricted tissue penetration of near-infrared light and is not suitable for precise visualization of deep tumor borders. In minimally invasive surgery, navigation is therefore primarily useful for localizing small lesions and visualizing essential anatomical structures such as the vasculature and biliary tree. These navigation systems typically rely on registration to preoperative imaging models, which provide a stable anatomical reference when intraoperative ultrasound is less feasible.

In this study, we generated US volumes by stacking tracked 2D US images. This approach eliminates the need for specialized 3D US probes, which are often bulky and can be difficult to maneuver in anatomically constrained regions (15). Nonetheless, even with conventional probes, imaging in areas with limited space (e.g., the liver dome) can be challenging and may result in deformation of the liver. Imaging very large or exophytic tumors presents an additional challenge, as maintaining stable and continuous probe contact during the sweep can be difficult. These factors may introduce artefacts that compromise image quality and reduce segmentation accuracy. In this feasibility study, ultrasound-based navigation was evaluated primarily in technically straightforward resections involving relatively small tumors, with a median diameter of 31 mm. Future work should focus on expanding the clinical cohort to demonstrate the broader benefits of this technology.

A remaining challenge in the current system is the visualization of navigation information during surgery. Surgeons reported difficulty in intuitively relating the position of instruments to the tumor anatomy using the current 2D screen layouts. In addition, maintaining continuous visual attention to the navigation screen was considered impractical during critical phases of the procedure. These findings highlight the need for improved visualization methods, as well as the integration of a warning trigger that alerts the surgical team when the instrument approaches the tumor margin.

The reference sensor plays a critical role in the navigation system by compensating for intraoperative liver motion. Particularly during resection of large tumors and tumors in more flexible segments, such as segments 2 and 6, accurate tracking requires the liver sensor to be positioned as close as possible to the intended resection plane. Incorrect placement can adversely affect the accuracy of navigation. In the current setup, sensor detachment occurred in 15% of patients during tumor resection, leading to complete loss of navigation. This detachment was caused by mechanical tension applied to the sensor cable. Future implementations using wireless electromagnetic sensors can mitigate this problem.

An important limitation of the study protocol relates to the accuracy assessment based on surgical clips placed along the resection plane. First, the surgical clips were 3.8 mm in length, and although the surgeon was instructed to point the center of each clip intraoperatively using a tracked pointer, this step is subject to some degree of uncertainty. Additionally, in some cases, clips detached from the specimen during handling or positioning in the pathology tray. For large specimens, it was critical to avoid placing the resection surface directly on the tray during the postoperative CT scan, as this could result in deformation of the specimen and artificially reduce the measured distance between the clips and the tumor. Finally, tissue shrinkage due to coagulation and loss of blood supply following resection may further affect the accuracy of distance measurements.

In conclusion, the proposed system utilizing intraoperative US imaging for navigation not only enhances accuracy but also allows its application throughout the resection process. The integration of automated segmentation further streamlines the workflow, supporting clinical implementation. While current visualization of the navigation should be improved, this study establishes a foundation for simpler and more accurate image-guided systems in liver surgery.

# References

1. Benedetti Cacciaguerra A, Dagher I, Fuks D, Rotellar F, D'Hondt M, Troisi RI, et al. Risk Factors of R1 Resection in Laparoscopic and Open Liver Surgery for Colorectal Liver Metastases: A European Multicentre Study. HPB [Internet]. 2021 Jan 1 [cited 2025 Jul 14];23:S18. Available from: https://www.hpbonline.org/action/showFullText?pii=S1365182X20312879
2. Görgec B, Benedetti Cacciaguerra A, Lanari J, Russolillo N, Cipriani F, Aghayan D, et al. Assessment of Textbook Outcome in Laparoscopic and Open Liver Surgery. JAMA Surg [Internet]. 2021 Aug 1 [cited 2025 Jul 14];156(8):e212064. Available from: https://pmc.ncbi.nlm.nih.gov/articles/PMC8173471/
3. Olthof K, Smit J, Fusaglia M, Kok N, Ruers T, Kuhlmann K. A surgical navigation system to aid the ablation of vanished colorectal liver metastases. Br J Surg [Internet]. 2024 May 3 [cited 2024 Dec 11];111(5). Available from: https://dx.doi.org/10.1093/bjs/znae110
4. Kingham TP, Pak LM, Simpson AL, Leung U, Doussot A, D'Angelica MI, et al. 3D image guidance assisted identification of colorectal cancer liver metastases not seen on intraoperative ultrasound: results from a prospective trial. Hpb [Internet]. 2018;20(3):260–7. Available from: http://dx.doi.org/10.1016/j.hpb.2017.08.035
5. Tinguely P, Fusaglia M, Freedman · Jacob, Banz V, Weber S, Candinas D, et al. Laparoscopic image-based navigation for microwave ablation of liver tumors-A multi-center study. Surg Endosc. 0.
6. Peterhans M, Vom Berg A, Dagon B, Inderbitzin D, Baur C, Candinas D, et al. A navigation system for open liver surgery: design, workflow and first clinical applications. Int J Med Robot Comput Assist Surg [Internet]. 2011 Mar 1 [cited 2025 Mar 25];7(1):7–16. Available from: https://onlinelibrary.wiley.com/doi/full/10.1002/rcs.360
7. Huber T, Tripke V, Baumgart J, Bartsch F, Schulze A, Weber S, et al. Computer-assisted intraoperative 3D-navigation for liver surgery: a prospective randomized-controlled pilot study. Ann Transl Med [Internet]. 2023 [cited 2025 Mar 25];11(10). Available from: https://dx.doi.org/10.21037/atm-22-5489
8. Banz VM, Müller PC, Tinguely P, Inderbitzin D, Ribes D, Peterhans M, et al. Intraoperative image-guided navigation system: development and applicability in 65 patients undergoing liver surgery. Langenbeck's Arch Surg. 2016;401(4):495–502.
9. Ivashchenko O V, Koert †, Kuhlmann FD, Van Veen R, Pouw B, Kok NFM, et al. CBCT-based navigation system for open liver surgery: Accurate guidance toward mobile and deformable targets with a semi-rigid organ approximation and electromagnetic tracking of the liver. 2021 [cited 2025 Mar 25]; Available from: https://doi.org/10.1002/mp.14825]
10. Smit JN, Kuhlmann KF, Ivashchenko O V, Thomson BR, Langø T, Kok NF, et al. Ultrasound-based navigation for open liver surgery using active liver tracking. Int J Comput Assist Radiol Surg. 2022;
11. Smit JN, Kuhlmann KFD, Thomson BR, Kok NFM, Ruers TJM, Fusaglia M. Ultrasound guidance in navigated liver surgery: toward deep-learning enhanced compensation of deformation and organ motion. Int J Comput Assist Radiol Surg [Internet]. 2024 Jan 1 [cited 2025 Mar 26];19(1):1–9. Available from: https://pubmed.ncbi.nlm.nih.gov/37249749/
12. Tu P, Hu P, Wang J, Chen X. From Coarse to Fine: Non-Rigid Sparse-Dense Registration for Deformation-Aware Liver Surgical Navigation. IEEE Trans Biomed Eng. 2024;71(9):2663–77.
13. Clements LW, Collins JA, Weis JA, Simpson AL, Kingham TP, Jarnagin WR, et al. Deformation Correction for Image-Guided Liver Surgery: An Intraoperative Assessment of Fidelity. Surgery [Internet]. 2017 Sep 1 [cited 2025 Mar 26];162(3):537. Available from: https://pmc.ncbi.nlm.nih.gov/articles/PMC5836469/

14. Lange T, Papenberg N, Heldmann S, Modersitzki J, Fischer B, Lamecker H, et al. 3D ultrasound-CT registration of the liver using combined landmark-intensity information. Int J Comput Assist Radiol Surg [Internet]. 2009 [cited 2025 Mar 26];4(1):79–88. Available from: https://pubmed.ncbi.nlm.nih.gov/20033605/
15. Beller S, Hünerbein M, Eulenstein S, Lange T, Schlag PM. Feasibility of navigated resection of liver tumors using multiplanar visualization of intraoperative 3-dimensional ultrasound data. Ann Surg [Internet]. 2007 Aug [cited 2025 Jun 6];246(2):288–94. Available from: https://pubmed.ncbi.nlm.nih.gov/17667508/
16. Chopra SS, Hünerbein M, Eulenstein S, Lange T, Schlag PM, Beller S. Development and validation of a three dimensional ultrasound based navigation system for tumor resection. Eur J Surg Oncol [Internet]. 2008 Apr [cited 2025 Jun 6];34(4):456–61. Available from: https://pubmed.ncbi.nlm.nih.gov/17765451/
17. Paolucci I, Sandu RM, Sahli L, Prevost GA, Storni F, Candinas D, et al. Ultrasound Based Planning and Navigation for Non-Anatomical Liver Resections – An Ex-Vivo Study . IEEE Open J Eng Med Biol. 2019;1:3–8.
18. Boretto L, Pelanis E, Regensburger A, Petkov K, Palomar R, Fretland ÅA, et al. Intraoperative patient-specific volumetric reconstruction and 3D visualization for laparoscopic liver surgery. Healthc Technol Lett. 2024 Dec 1;
19. Boretto L, Pelanis E, Regensburger A, Fretland ÅA, Edwin B, Elle OJ. Hybrid optical-vision tracking in laparoscopy: accuracy of navigation and ultrasound reconstruction. Minim Invasive Ther Allied Technol [Internet]. 2024 [cited 2025 Mar 25];33(3):176–83. Available from: https://www.tandfonline.com/doi/abs/10.1080/13645706.2024.2313032
20. Fedorov A, Beichel R, Kalpathy-Cramer J, Finet J, Fillion-Robin JC, Pujol S, et al. 3D Slicer as an image computing platform for the Quantitative Imaging Network. Magn Reson Imaging. 2012 Nov;30(9):1323–41.
21. Lasso A, Heffter T, Rankin A, Pinter C, Ungi T, Fichtinger G. PLUS: open-source toolkit for ultrasound-guided intervention systems. IEEE Trans Biomed Eng [Internet]. 2014 Oct 1 [cited 2025 Jul 11];61(10):2527. Available from: https://pmc.ncbi.nlm.nih.gov/articles/PMC4437531/
22. Pérez de Frutos J, Hofstad EF, Solberg OV, Tangen GA, Lindseth F, Langø T, et al. Laboratory test of Single Landmark registration method for ultrasound-based navigation in laparoscopy using an open-source platform. Int J Comput Assist Radiol Surg [Internet]. 2018 Dec 1 [cited 2024 Dec 12];13(12):1927–36. Available from: https://link.springer.com/article/10.1007/s11548-018-1830-7
23. Thomson BR, Nijkamp J, Ivashchenko O, van der Heijden F, Smit JN, Kok NFM, et al. Hepatic vessel segmentation using a reduced filter 3D U-Net in ultrasound imaging. 2019; Available from: http://arxiv.org/abs/1907.12109
24. T. Natali, K.A. Olthof, K.F.D. Kuhlmann, T.J.M. Ruers MF. Intraoperative assessment of automatic hepatic tumor segmentation for ultrasound navigation. ArXiv. 2025;
25. Achterberg FB, Bijlstra OD, Slooter MD, Sibinga Mulder BG, Boonstra MC, Bouwense SA, et al. ICG-Fluorescence Imaging for Margin Assessment During Minimally Invasive Colorectal Liver Metastasis Resection + Invited Commentary + Supplemental content. JAMA Netw Open. 2024;7(4):246548.